\documentclass[12pt]{article}

\usepackage[dvips]{graphicx}
\usepackage{amsmath}
\usepackage{amssymb}

\newcommand{\bnll}[1]{\begin{subequations}\label{#1}\begin{eqnarray}}
\newcommand{\enll}{\end{eqnarray}\end{subequations}}
\newcommand{\thalf}{\tfrac{1}{2}}

\let\section=\subsection
\let\subsection=\subsubsection

\newcounter{mysubsubsection}

\newcommand{\codeversion}{(v1.66p)}

\newcommand{\tbd}[1]{\noindent{\bf To be done: #1.}}

\begin{document}

\begin{center}

\vspace{0.5cm}
{\bf\Large
Axially Deformed Solution of the
Skyrme-Hartree-Fock-Bogolyubov Equations
using The Transformed Harmonic Oscillator Basis. \\[1ex]
The program HFBTHO \codeversion
}

\vspace{5mm} {\large M.V.~Stoitsov$^{\mathrm{a-d,}}$\footnote{Corresponding
author. {\it E-mail address:} stoitsovmv@ornl.gov.},
J.~Dobaczewski$^{\mathrm{a-c,e}}$,
W.~Nazarewicz$^{\mathrm{a,b,e}}$,
P.~Ring$^{\mathrm{f}}$ }

\vspace{3mm}
{\it
$^{\mathrm{b}}$Department of Physics and Astronomy, University
of Tennessee, Knoxville, TN 37996, USA  \\
\vspace{1mm}
$^{\mathrm{a}}$Physics Division,  Oak Ridge National
Laboratory, P.O.B.
2008, Oak Ridge, TN 37831, USA  \\
\vspace{1mm} $^{\mathrm{c}}$Joint Institute for Heavy Ion Research, Oak
Ridge, TN 37831, USA  \\
\vspace{1mm} $^{\mathrm{d}}$Institute of Nuclear Research  and Nuclear
Energy,  Bulgarian Academy of Sciences, \\ Sofia, Bulgaria \\
\vspace{1mm} $^{\mathrm{e}}$Institute of Theoretical Physics, Warsaw
University, ul.Ho\.za 69, 00-681 Warsaw, Poland  \\
\vspace{1mm} $^{\mathrm{f}}$Physics Department, Technical University Munich,
Garching, Germany \\
}

\end{center}

\vspace{5mm}

\hrule \vspace{5mm} \noindent{\bf Abstract}

We describe the program HFBTHO for axially deformed configurational
Hartree-Fock-Bogoliubov calculations with Skyrme-forces and zero-range
pairing interaction  using Harmonic-Oscillator  and/or Transformed
Harmonic-Oscillator states. The particle-number
symmetry is approximately restored
using the Lipkin-Nogami
prescription, followed by an exact particle number projection after the
variation.
The program  can be used  in a variety of applications, including
systematic studies of wide ranges of nuclei, both spherical and
axially deformed, extending all the way out to nucleon drip lines.

\vspace{5mm}
\noindent{\bf Program Summary}

\bigskip\noindent{\it Title of the program:} HFBTHO \codeversion

\bigskip\noindent{\it Catalogue number:}

\bigskip\noindent{\it Program obtainable from:}
                      CPC Program Library, \
                      Queen's University of Belfast, N. Ireland

\bigskip\noindent{\it Program summary URL:}

\bigskip\noindent{\it Licensing provisions:} none

\bigskip\noindent{\it Computers on which the program has been tested:}
                      Pentium-III, Pentium-IV, AMD-Athlon, IBM Power 3, IBM Power 4, Intel Xeon

\bigskip\noindent{\it Operating systems:} LINUX, Windows

\bigskip\noindent{\it Programming language used:} FORTRAN-95

\bigskip\noindent{\it Memory required to execute with typical data:} 59 MB  when using $N_{sh}=20$

\bigskip\noindent{\it No.\ of bits in a word:} 64

\bigskip\noindent{\it No.\ of processors used:} 1

\bigskip\noindent{\it Has the code been vectorized?:} No

\bigskip\noindent{\it No.\ of bytes in distributed program, including test data, etc.:}

\bigskip\noindent{\it No.\ of lines in distributed program:} 7876 lines

\bigskip\noindent{\it Nature of physical problem:}
Solution of self-consistent mean-field
equations for weakly bound paired nuclei requires
correct description of asymptotic properties of nuclear quasiparticle
wave functions. In the present implementation, this is achieved
 by using the single-particle
wave functions of the
Transformed Harmonic Oscillator, which allows for an
accurate description of  deformation effects and pairing correlations
in nuclei arbitrarily close to the particle drip lines.

\bigskip\noindent{\it Method of solution:}
The program uses axial Transformed Harmonic Oscillator single-particle
basis to expand quasiparticle wave functions. It  iteratively
diagonalizes the Hartree-Fock-Bogolyubov Hamiltonian
based on the Skyrme forces and zero-range pairing interaction until the
self-consistent solution is achieved.

\bigskip\noindent{\it Restrictions on the complexity of the problem:}
Axial-, time-reversal-, and space-inversion symmetries are assumed.
Only quasiparticle vacua  of even-even nuclei can be calculated.

\bigskip\noindent{\it Typical running time:} 4 seconds per iteration on an Intel Xeon 2.8 GHz processor when using $N_{sh}=20$

\bigskip\noindent{\it Unusual features of the program:} none

\vspace{5mm}



\bigskip

\section{Introduction}
\label{sec0}

Nuclear structure theory strives to build a
comprehensive microscopic framework in which bulk nuclear properties,
nuclear excitations, and
nuclear reactions can all be described. Exotic
radioactive nuclei are the critical new focus in this quest. The
extreme isospin of these nuclei and their weak binding bring new
phenomena that amplify important features of the
nuclear many-body problem.

A proper theoretical description of such weakly bound systems
requires a careful treatment of the asymptotic part of the
nucleonic density. An appropriate framework for these calculations
is Hartree-Fock-Bogoliubov (HFB) theory, solved in coordinate
representation \cite{[Bul80],[Dob84]}. This method has
been used extensively in the treatment of spherical nuclei
\cite{[Dob96]}  but is
much more difficult to implement for systems with deformed
equilibrium shapes. There have been three ways of implementing
deformation effects into the coordinate-space HFB. The oldest
method, the so-called two-basis method \cite{[Gal94],[Ter97a],[Yam01]},
is based on the diagonalization of the particle-particle
part of the HFB Hamiltonian in the self-consistent  basis,
obtained by solving the HF problem with box boundary conditions.
The disadvantage of this method is the appearance of a large number
of positive-energy free-particle (box) states, which  limits
the number of discretized continuum states
(the maximum
single-particle energy taken in this
method is usually less than 10\,MeV).

The second, very promising
 strategy, the so-called canonical-basis HFB
method, utilizes the spatially localized eigenstates of the one-body
density matrix without explicitly going to the quasiparticle
representation \cite{[Rei97],[Taj98],[Taj04]}. Finally, an approach to
axial coordinate-space HFB has recently been developed  that uses
a basis-spline method \cite{[Ter03],[Obe03]}. While precise, these
two latter methods are not easy to implement and,
because they are time-consuming, cannot be used in large-scale
calculations in which a crucial factor is the ability to
perform quick calculations for many nuclei.

In the absence of fast  coordinate-space solutions to the
deformed HFB equations, it is useful to consider instead the
configuration-space approach, whereby the HFB solution is expanded
in some single-particle basis. In this context, the basis of a harmonic
oscillator (HO) turned out to be particularly useful. Over the years, many
configuration-space HFB+HO codes have  been developed,
either employing Skyrme forces or the Gogny
effective interaction \cite {[Gog75],[Gir83],[Egi80],[Egi95],[Dob04]}, or
using a relativistic Lagrangian \cite{[Rin96]} in the context
of the relativistic Hartree-Bogoliubov theory. For nuclei at the
drip lines, however, the HFB+HO expansion converges slowly as a
function of the number of oscillator shells \cite{[Dob96]},
producing wave functions that decay too rapidly at large
distances.

A related alternative approach that has recently been proposed is to
expand the quasiparticle HFB wave functions in a complete set of
transformed harmonic oscillator (THO) basis states
\cite{[Sto98b]}, obtained by applying a
local-scaling coordinate transformation (LST)
\cite{[Sto83],[Sto91]} to the standard HO basis. Applications of
this HFB+THO methodology have been reported both in the
non-relativistic \cite{[Sto99]} and relativistic
domains \cite{[Sto98c]}. In all of these calculations, specific
global parameterizations were employed for the scalar LST function
that defines the THO basis. There are several limitations in such
an approach, however. For example, the minimization procedure that
is needed in such an approach to optimally define the basis
parameters is computationally very time-consuming, making it very
difficult to apply the method systematically to nuclei across the
periodic table.

Recently, a new prescription for choosing the THO basis has been
proposed and employed in self-consistent large-scale
calculations \cite{[Sto03]}. For a given nucleus, the new prescription
requires as input the results from a relatively simple HFB+HO
calculation, with no variational optimization. The resulting THO
basis leads to HFB+THO results that almost exactly reproduce the
coordinate-space HFB results for spherical nuclei \cite{[Dob04a]}.
Because the new prescription requires no variational optimization
of the LST function, it can be applied in systematic studies of
nuclear properties. In order to correct for the particle number
nonconservation inherent to the HFB approach,
 the Lipkin-Nogami
prescription for an approximate particle number projection,
followed by an exact particle number projection after the
variation has been implemented the code HFBTHO {\codeversion} \cite{[Sto04a],[Sto04b]}.

The paper is organized as follows.
Section~\ref{sec1} gives a brief summary of the HFB formalism. The implementation
of the method to the case of the Skyrme energy density functional is discussed
in Sec.~\ref{sec2}, together with the overview of the THO method and the
treatment of pairing. Section~\ref{sec3} describes
the code HFBTHO {\codeversion}.
Finally, conclusions are given in Sec.~\ref{sec4}.


\section{Hartree-Fock-Bogoliubov Method}
\label{sec1}

A two-body Hamiltonian of a system of fermions can
be expressed in terms of a set of annihilation and creation
operators $(c,c^{\dagger })$:
\begin{eqnarray}
H &=&\sum_{n_{1}n_{2}}e_{n_{1}n_{2}}~c_{n_{1}}^{\dagger }c_{n_{2}}
 +\tfrac{1}{4}\sum_{n_{1}n_{2}n_{3}n_{4}}\overline{v}
_{n_{1}n_{2}n_{3}n_{4}}~c_{n_{1}}^{\dagger }c_{n_{2}}^{\dagger
}c_{n_{4}}c_{n_{3}},  \label{E1}
\end{eqnarray}
where $ \overline{v}_{n_{1}n_{2}n_{3}n_{4}}=\langle
n_{1}n_{2}|V|n_{3}n_{4}-n_{4}n_{3}\rangle$ are anti-symmetrized
two-body interaction matrix-elements. In the HFB method, the ground-state
wave function $|\Phi \rangle $ is defined as the quasiparticle vacuum
$\alpha _{k}|\Phi \rangle =0$, where the quasiparticle operators
$(\alpha ,\alpha^{\dagger })$ are connected to the original
particle operators via the linear Bogoliubov transformation
\begin{eqnarray}
\quad \quad \alpha_{k} &=&\sum_{n}\left( U_{nk}^{\ast
}c_{n}+V_{nk}^{\ast }c_{n}^{\dagger }\right) ,  \label{E4}  \quad
\quad \alpha_{k}^{\dagger } =\sum_{n}\left(
V_{nk}c_{n}+U_{nk}c_{n}^{\dagger }\right) ,  \label{E5}
\end{eqnarray}
which can be rewritten in the matrix form as
\begin{equation}
\left(
\begin{array}{c}
\alpha \\
\alpha^{\dagger }
\end{array}
\right) =\left(
\begin{array}{cc}
U^{\dagger } & V^{\dagger } \\
V^{T} & U^{T}
\end{array}
\right) \left(
\begin{array}{c}
c \\
c^{\dagger }
\end{array}
\right)  .  \label{E6}
\end{equation}
Matrices $U$ and $V$ satisfy the relations:
\begin{equation}
U^{\dagger }U+V^{\dagger }V=I,~~~UU^{\dagger }+V^{\ast
}V^{T}=I,~~~U^{T}V+V^{T}U=0,~~~UV^{\dagger }+V^{\ast }U^{T}=0.
\label{E9}
\end{equation}
In terms of the normal $\rho $ and pairing $\kappa $ one-body
density matrices, defined as
\begin{equation}
\rho_{nn^{\prime }}~=~\langle \Phi |c_{n^{\prime }}^{\dagger
}c_{n}|\Phi \rangle=(V^{\ast }V^{T})_{nn^{\prime }} ,\quad
\kappa_{nn^{\prime }}~=~\langle \Phi |c_{n^{\prime }}c_{n}|\Phi
\rangle=(V^{\ast }U^{T})_{nn^{\prime }} , \label{E11}
\end{equation}
the expectation value of the Hamiltonian (\ref{E1}) is expressed
as an energy functional
\begin{eqnarray}
E[\rho ,\kappa ] &=&\frac{\langle \Phi |H|\Phi \rangle }{\langle
\Phi |\Phi \rangle }  ={\rm Tr}\left[ (e+\tfrac{1}{2}\Gamma )\rho
\right] -\tfrac{1}{2}{\rm Tr} \left[ \Delta \kappa^{\ast }\right] ,
\label{ehfb}
\end{eqnarray}
where
\begin{eqnarray}
\Gamma_{n_{1}n_{3}}
&=&\sum_{n_{2}n_{4}}\overline{v}_{n_{1}n_{2}n_{3}n_{4}}
\rho_{n_{4}n_{2}},~~~  \label{E102}  \Delta_{n_{1}n_{2}}
=\tfrac{1}{2}\sum_{n_{3}n_{4}}\overline{v}
_{n_{1}n_{2}n_{3}n_{4}}\kappa_{n_{3}n_{4}}.  \label{E103}
\end{eqnarray}
Variation of energy (\ref{ehfb}) with respect to  $\rho $ and
$\kappa $  results in the HFB equations:
\begin{equation}  \label{hfb}
\left(
\begin{array}{cc}
e+\Gamma -\lambda & \Delta \\
-\Delta^{\ast } & -(e+\Gamma )^{\ast }+\lambda
\end{array}
\right) \left(
\begin{array}{c}
U \\
V
\end{array}
\right) =E\left(
\begin{array}{c}
U \\
V
\end{array}
\right) ,
\end{equation}
where the Lagrange multiplier $\lambda $ has been  introduced to fix
the correct average particle number.

It should be stressed that the modern energy functionals
(\ref{ehfb}) contain terms that cannot be simply related
to  some prescribed  effective interaction, see e.g., Ref.\
\cite{[Ben03],[Per04]}
for details. In this respect the
functional (\ref{ehfb}) should be considered in the broader
context of the energy density functional theory.


\section{Skyrme Hartree-Fock-Bogoliubov Method}
\label{sec2}

\subsection{Skyrme Energy Density Functional}
\label{sec21}

For Skyrme forces, the HFB energy  (\ref{ehfb}) has the form of
a local energy density functional,
\begin{equation}
E[\rho,\tilde{\rho}]=\int d^3{\bf r}~{\cal H}({\bf r}) , \label{shfb}
\end{equation}
where
\begin{equation}
{\cal H}({\bf r})=H({\bf r})+\tilde{H}({\bf r})
\label{enden}
\end{equation}
is the sum of the
mean-field and pairing energy densities. In the present implementation,
we use the following explicit forms:
\begin{equation}
\begin{array}{rll}
H({\bf r}) & = & \tfrac{\hbar^{2}}{2m}\tau +\tfrac{1}{2}t_{0}\left[~
\left( 1+\tfrac{1}{2}x_{0}\right) \rho^{2}\right. -\left(
\tfrac{1}{2}+~x_{0}\right) \sum\limits_{q}\left.
\rho_{q}^{2}~ \right] \\
& + & \tfrac{1}{2}t_{1}\left[~ \left( 1+\tfrac{1}{2}x_{1}\right)
\rho \left( \tau -\tfrac{3}{4}\left. \Delta \rho \right) \right.
\right] -\left(\tfrac{1}{2}+~x_{1}\right) \sum\limits_{q}\left.
\rho
_{q}\left( \tau_{q}-\tfrac{3}{4}\Delta \rho_{q}\right) \right] \\
& + & \tfrac{1}{2}t_{2}\left[~ \left( 1+\tfrac{1}{2}x_{2}\right)
\rho \left( \tau +\tfrac{1}{4}\Delta \rho \right) \right. -\left(
\tfrac{1}{2}+~x_{2}\right) \sum\limits_{q}\left. \rho
_{q}\left( \tau_{q}+\tfrac{1}{4}\Delta \rho_{q}\right) \right] \\
& + & \tfrac{1}{12}t_{3}\rho^{\alpha }\left[ \left( 1+\tfrac{1}{2}
x_{3}\right) \rho^{2}\right. -\left( x_{3}+\tfrac{1}{2}\right)
\sum\limits_{q}\left. \rho_{q}^{2}\right] \\
& - & \tfrac{1}{8}\left( t_{1}x_{1}+t_{2}x_{2}\right)
\sum\limits_{ij}{\bf J} _{ij}^{2}   +  \tfrac{1}{8}\left(
t_{1}-t_{2}\right) \sum\limits_{q,ij}{\bf J}
_{q,ij}^{2} \\
& - & \tfrac{1}{2}W_{0}\sum\limits_{ijk}\varepsilon_{ijk}\left[
\rho {\bf \nabla }_{k}{\bf J}_{ij}\right. +\sum\limits_{q}\left.
\rho_{q}{\bf \nabla }_{k}{\bf J}_{q,ij}\right]
\end{array}
\label{skyrmeph}
\end{equation}
and
\begin{equation}
\displaystyle \tilde{H}({\bf r}) = \tfrac{1}{2}V_0
\left[1-V_1\left(\frac{\rho}{\rho_0}\right)^\gamma~
\right]\sum\limits_{q}\tilde{\rho}_{q}^{2} . \label{edhppd}
\end{equation}
Index $q$ labels the neutron ($q=n$) or proton
($q=p$) densities, while densities without index $q$ denote
the sums of proton and neutron densities.
$H({\bf r})$ and $\tilde{H}({\bf r})$ depend on the
particle local density $\rho ({\bf r})$, pairing local density
$\tilde{\rho}({\bf r})$, kinetic energy density $ \tau ({\bf
r})$,  and spin-current density ${\bf
J}_{ij}({\bf r})$:
\begin{equation}
\begin{array}{c}
\begin{array}{ccl}
\rho ({\bf r}) & = & \rho ({\bf r},{\bf r}), \\
~ &  &  \\
\tau ({\bf r}) & = & \left. \nabla_{{\bf r}}\nabla_{{\bf
r}^{\prime }}\rho
({\bf r},{\bf r}^{\prime })\right|_{{\bf r}^{\prime }{\bf =r}}\;, \\
\end{array}
\begin{array}{ccl}
\tilde{\rho}({\bf r}) & = & \tilde{\rho}({\bf r},{\bf r}), \\
~ &  &  \\
{\bf J}_{ij}({\bf r}) & = & \tfrac{1}{2i}\left. \left( \nabla
_{i}-\nabla_{i}^{\prime }\right) \rho_{j}({\bf r},{\bf r}^{\prime
})\right|_{{\bf r}^{\prime }{\bf =r}}\;,
\end{array}
\end{array}
\label{densities}
\end{equation}
where $\rho ({\bf r},{\bf r}^{\prime }),\,\rho_{i}({\bf r},{\bf
r}^{\prime }),\,\tilde{\rho}({\bf r},{\bf r}^{\prime
}),\,\tilde{\rho}_{i}({\bf r},{\bf r}^{\prime })$ are defined by
the spin-dependent one-body density matrices in the standard way:
\begin{equation}
\begin{array}{rcl}
\rho ({\bf r}\sigma ,{\bf r^{\prime }}\sigma^{\prime })
&=&\displaystyle\tfrac{1}{2} \rho ({\bf r},{\bf r}^{\prime })\delta_{\sigma
\sigma^{\prime }} + \tfrac{1}{2} \sum_{i}(\sigma
|\sigma_{i}|\sigma^{\prime })\rho_{i}({\bf r},{\bf r}
^{\prime }) ,   \\ && \\
\tilde{\rho}({\bf r}\sigma ,{\bf r^{\prime }}\sigma^{\prime })
&=&\displaystyle\tfrac{1}{2 } \tilde{\rho}({\bf r},{\bf r}^{\prime })\delta
_{\sigma \sigma^{\prime }}+\tfrac{1}{2} \sum_{i}(\sigma
|\sigma_{i}|\sigma^{\prime })\tilde{\rho}_{i}( {\bf r},{\bf
r}^{\prime }) .
\label{densitieemp}
\end{array}
\end{equation}
We use the pairing density matrix $\tilde{\rho}$,
\begin{equation}
\tilde{\rho}({\bf r}\sigma ,{\bf r^{\prime }}\sigma^{\prime
})=-2\sigma^{\prime }\kappa ({\bf r,}\sigma ,{\bf r^{\prime
},}-\sigma^{\prime }) ,
\end{equation}
instead of the pairing tensor $\kappa $. This is convenient when
describing time-even quasiparticle states when both $\rho $
and $\tilde{\rho}$ are hermitian and time-even \cite{[Dob84]}.
In the pairing energy density (\ref{edhppd}), we have restricted our
consideration to contact delta pairing forces in order to reduce the
complexity of the general expressions \cite{[Dob84],[Per04]}.

\subsection{Skyrme Hartree-Fock-Bogoliubov Equations}
\label{sec22}

Variation of the energy (\ref{shfb}) with respect to  $\rho $ and
$\tilde{\rho} $  results in the Skyrme HFB equations:
\begin{equation}
\label{eq143}
\sum_{\sigma'}
\left(\begin{array}{cc}
h({\bf
r},\sigma,\sigma') &
\tilde h({\bf r},\sigma,\sigma') \\
\tilde h({\bf r},\sigma,\sigma') &
- h({\bf r},\sigma,\sigma')
\end{array}\right)
\left(\begin{array}{c}
U (E,{\bf r}\sigma') \\
V (E,{\bf r}\sigma')
\end{array}\right) =
\left(\begin{array}{cc}
E+\lambda & 0         \\
0    & E-\lambda \end{array}\right)
\left(\begin{array}{c}
U (E,{\bf r}\sigma) \\
V (E,{\bf r}\sigma)
\end{array}\right),
\end{equation}
where local fields $ h({\bf r},\sigma ,\sigma^{\prime })$ and
$\tilde{h}({\bf r},\sigma ,\sigma^{\prime })$ can be easily
calculated in the coordinate space by using the following explicit
expressions:
\begin{equation}
\begin{array}{rcl}
h_{q}({\bf r},\sigma ,\sigma^{\prime }) &=&-{\bf \nabla
}M_{q} {\bf \nabla } +U_{q} +\tfrac{1}{2i}\sum\limits_{ij}\left(
{\bf \nabla }_{i}\sigma _{j}B_{q,ij}+B_{q,ij}{\bf \nabla
}_{i}\sigma_{j}\right) , \\
\tilde{h}_{q}({\bf r},\sigma ,\sigma^{\prime }) &=& \displaystyle
 V_0 \left(1-V_1 \left(\frac{\rho}{\rho_0}\right)^\gamma \right)\tilde{\rho}_{q},
\label{hppq}
\end{array}
\end{equation}
where
\begin{equation}
\begin{array}{lcl}
M_{q} &=&\frac{\hbar^{2}}{2m}+\tfrac{1}{4}t_{1}\left[ \left(
1+\tfrac{1}{2} x_{1}\right) \rho -\left( x_{1}+\tfrac{1}{2}\right)
\rho_{q}^{2}\right]  +\tfrac{1}{4}t_{2}\left[ \left(
1+\tfrac{1}{2}x_{2}\right) \rho +\left( x_{2}+\tfrac{1}{2}\right)
\rho_{q}^{2}\right] ,
\\ &~& \\
B_{q,ij} &=&-\tfrac{1}{4}\left( t_{1}x_{1}+t_{2}x_{2}\right) {\bf
J}_{ij} +\tfrac{1}{4}\left( t_{1}-t_{2}\right) {\bf J}_{q,ij}
+\tfrac{1}{2}W_{0}\sum\limits_{ijk}\varepsilon_{ijk}{\bf \nabla }
_{k}\left( {\bf \rho }+{\bf \rho }_{q},\right), \\
U_{q} &=&t_{0}\left[ \left( 1+\tfrac{1}{2}x_{0}\right) \rho -\left(
x_{0}+ \tfrac{1}{2}\right) \rho_{q}\right]
\\ &~& \\
&+&\tfrac{1}{4}t_{1}\left[ ~\left( 1+\tfrac{1}{2}x_{1}\right) \left(
\tau - \tfrac{3}{2}\Delta \rho \right) \right.  -\left(
x_{1}+\tfrac{1}{2}\right) \left. \left( \tau _{q}-\tfrac{3}{2}
\Delta \rho_{q}\right) \right]
\\ &~& \\
&+&\tfrac{1}{4}t_{2}\left[ ~\left( 1+\tfrac{1}{2}x_{2}\right) \left(
\tau + \tfrac{1}{2}\Delta \rho \right) \right. +\left(
x_{2}+\tfrac{1}{2}\right) \left. \left( \tau _{q}+\tfrac{1}{2}
\Delta \rho_{q}\right) \right]
\\ &~& \\
&+&\tfrac{1}{12}t_{3}\rho^{\alpha }\left[ \left(
1+\tfrac{1}{2}x_{3}\right) (2+\alpha )\rho \right.  -2\left(
x_{3}+\tfrac{1}{2}\right) \rho_{q} +\left( 1-x_{3}\right)
\frac{\alpha }{\rho }\left. \right]  -
\frac{\gamma V_0 V_1}{2 \rho} \left( \frac{\rho}{\rho_0} \right)^\gamma \sum\limits_{q}\tilde{\rho}_{q}^{2}
\\ &~& \\
&-&\tfrac{1}{8}\left( t_{1}x_{1}+t_{2}x_{2}\right)
\sum\limits_{ij}{\bf J} _{ij}^{2}  +\tfrac{1}{8}\left(
t_{1}-t_{2}\right) \sum\limits_{q,ij}{\bf J}_{q,ij}^{2}
-\tfrac{1}{2}W_{0}\sum\limits_{ijk}\varepsilon_{ijk}{\bf \nabla
}_{k}\left[ {\bf J}_{ij}\right. +\left. {\bf J}_{q,ij}\right] .
\end{array}
\label{mt}
\end{equation}

Properties of the HFB equation in the spatial coordinates, Eq.{\
}(\ref{eq143}), have been discussed in Ref.{\ }\cite{[Dob84]}.  In
particular, it has been shown that the spectrum of eigenenergies
$E$ is continuous for $|E|$$>$$-\lambda$ and discrete for
$|E|$$<$$-\lambda$.
In the present implementation, we solve the HFB equations
by expanding quasiparticle wave functions on a finite basis;
therefore, the quasiparticle spectrum $E_k$ becomes discretized.
Hence in the following we use the notation
$V_k({\bf r}\sigma)=V(E_k,{\bf r}\sigma)$ and $U_k({\bf r}\sigma)
=U(E_k,{\bf r}\sigma)$.
Since for $E_k$$>$0 and $\lambda$$<$0 the lower
components $V_k({\bf r}\sigma)$ are localized functions of ${\bf
r}$, the density matrices,
\begin{eqnarray}
\rho({\bf r}\sigma,{\bf r}'\sigma') &=& ~~ \sum_k  V_k
({\bf r} \sigma ) V_k^*({\bf r}'\sigma')
,   \label{eq144a} \\
\tilde\rho({\bf r}\sigma,{\bf r}'\sigma') &=&
- \sum_k  V_k({\bf r} \sigma ) U_k^*({\bf
r}'\sigma') ,   \label{eq144b}
\end{eqnarray}
are always localized. The orthogonality relation for the
single-quasiparticle HFB wave functions reads
\begin{equation}\label{orthog}
\int\text{d}^3{\bf r} \sum_{\sigma} \left[ U_k^*({\bf r} \sigma
) U_{k'}  ({\bf r}\sigma) + V_k^*({\bf r} \sigma ) V_{k'}
({\bf r}\sigma) \right] = \delta_{k,k'} ,
\end{equation}
and the norms of lower components $N_k$,
\begin{equation}\label{norms}
N_k  =  \int\text{d}^3{\bf r}\sum_{\sigma}|V_k({\bf r} \sigma
)|^2,
\end{equation}
define the total number of particles
\begin{equation}
\label{Ntot}
N=\int\text{d}^3{\bf r}~\rho({\bf r}) = \sum_{n}N_k.
\end{equation}

\subsection{Axially Deformed Nuclei}
\label{sec24}

For spherical nuclei, Skyrme HFB equations are best solved in the
coordinate space, because  Eq.~(\ref{eq143}) reduces in this case to a
set of radial differential equations \cite{[Ben04]}. In the case of
deformed nuclei, however, the solution of a deformed HFB equation in
coordinate space is a difficult and time-consuming task. For this
reason, here we use the method proposed by Vautherin \cite{[Vau73]},
which combines two different representations. The solution of the
deformed HFB equation is carried out by diagonalizing the HFB
hamiltonian in the configurational space of wave-functions with
appropriate symmetry, while evaluation of the potentials and
densities is performed in the coordinate space. Such a method is
applicable to nonaxial deformations \cite{[Dob04]}, but typical
computation time for large-scale mass-table calculations is
prohibitively large. In the present implementation, we make the
restriction to axially-symmetric and reflection-symmetric shapes in order to
obtain HFB solutions within a much shorter CPU time.

In the case of axial symmetry, the third
component $J_z$ of the total angular momentum is conserved and
provides a good quantum number $\Omega_k$.
Therefore, quasiparticle HFB states can be written in the following form:
\begin{equation}
\left(
\begin{array}{c}
U_k ({\bf r},\sigma,\tau) \\
V_k ({\bf r},\sigma,\tau)
\end{array}\right)
= \chi_{q_k}(\tau) \left[
\left(
\begin{array}{c}
U_k^+ (r,z) \\
V_k^+ (r,z)
\end{array}\right)
e^{\imath \Lambda^- \varphi} \chi_{+1/2}(\sigma) + \left(
\begin{array}{c}
U_k^- (r,z) \\
V_k^- (r,z)
\end{array}\right)
e^{\imath \Lambda^+ \varphi} \chi_{-1/2}(\sigma) \right],
\label{hfbs}
\end{equation}
where $\Lambda^\pm=\Omega_k \pm 1/2$ and $r$, $z$, and $\varphi$
are the standard cylindrical coordinates defining the three-dimensional
position vector as ${\bf r}=(r\cos\varphi,r\sin\varphi,z)$,
while $z$ is the chosen symmetry
axis. The quasiparticle states (\ref{hfbs}) are also assumed to be
eigenstates of the third component of the isospin operator with
eigenvalues $q_k=+1/2$ for protons and $q_k=-1/2$ for neutrons.

By substituting ansatz (\ref{hfbs}) into Eq.~(\ref{eq143}),
the HFB equation reduces to a system of equations involving the
cylindrical variables $r$ and $z$ only. The same is also true for the
local densities, i.e.,
\begin{equation}
\begin{array}{rcll}
\rho (r,z) & = & \displaystyle\sum_{k} &\displaystyle \left(
|V_{k}^{+}(r,z)|^{2}+|V_{k}^{-}(r,z)|^{2}\right),  \\
\tau (r,z) & = & \displaystyle\sum_{k} &\displaystyle \left( |\nabla
_{r}V_{k}^{+}(r,z)|^{2}+|\nabla _{r}V_{k}^{-}(r,z)|^{2}+\frac{1}{r^{2}}
|\Lambda ^{+}V_{k}^{-}(r,z)|^{2}\right.  \\
&  &  &\displaystyle + \left. |\nabla _{z}V_{k}^{+}(r,z)|^{2}+|\nabla
_{z}V_{k}^{-}(r,z)|^{2}+\frac{1}{r^{2}}|\Lambda
^{-}V_{k}^{+}(r,z)|^{2}\right),  \\
(\mathbf{\nabla}\cdot\mathbf{J})(r,z) & = & \displaystyle\sum_{k} & \displaystyle\left(
\nabla_{r}V_{k}^{+}(r,z)\nabla _{z}{V_{k}^{-}}(r,z)
+\frac{\Lambda ^{-}}{r}V_{k}^{+}(r,z)\left[ \nabla
_{r}V_{k}^{+}(r,z)-\nabla _{z}{V_{k}^{-}}(r,z)\right] \right. \\
& & &\displaystyle \left. -\nabla_{r}V_{k}^{-}(r,z)\nabla _{z}{V_{k}^{+}}(r,z)
-\frac{\Lambda ^{+}}{r}V_{k}^{-}(r,z)\left[ \nabla
_{r}V_{k}^{-}(r,z)+\nabla _{z}{V_{k}^{+}}(r,z)\right] \right) \\
\tilde{\rho}(r,z) & = & - \displaystyle\sum_{k} &\displaystyle
\left( V_{k}^{+}(r,z){U_{k}^{+}}(r,z)+V_{k}^{-}(r,z){U_{k}^{-}}(r,z)\right),
\end{array}
\label{locden}
\end{equation}
where $\nabla_r=\partial/\partial r$ and $\nabla_z=\partial/\partial z$.
In addition, when tensor forces are considered, the following additional densities
have to be calculated:
\begin{equation}
\begin{array}{rcll}
J_{r \varphi} (r,z) & = & \displaystyle\sum_{k} &\displaystyle \left(
\nabla_{r}V_{k}^{+}(r,z)V_{k}^{-}(r,z) - \nabla_{r}V_{k}^{-}(r,z)V_{k}^{+}(r,z) \right), \\
J_{\varphi r} (r,z) & =& \displaystyle\sum_{k} &\displaystyle \left(
\frac{\Lambda^{-}}{r} V_{k}^{+}(r,z)V_{k}^{-}(r,z)+
\frac{\Lambda^{+}}{r} V_{k}^{-}(r,z)V_{k}^{+}(r,z) \right),\\
J_{z \varphi} (r,z) & = & \displaystyle\sum_{k} &\displaystyle \left(
\nabla_{z}V_{k}^{+}(r,z)V_{k}^{-}(r,z) - \nabla_{z}V_{k}^{-}(r,z)V_{k}^{+}(r,z) \right), \\
J_{\varphi z} (r,z) & = & \displaystyle\sum_{k} &\displaystyle \left(
\frac{\Lambda^{-}}{r} V_{k}^{+}(r,z)V_{k}^{+}(r,z)-
\frac{\Lambda^{+}}{r} V_{k}^{-}(r,z)V_{k}^{-}(r,z) \right),\\
\end{array}
\label{locten}
\end{equation}
where indices denote the cylindrical components of the tensor ${\bf J}_{ij}$,
while all remaining components vanish due to the cylindrical symmetry,
i.e., $J_{r r}(r,z) = J_{z z}(r,z) = J_{\varphi\varphi }(r,z) = J_{r z}(r,z) = J_{z r}(r,z)$=0

Due to the time-reversal symmetry, if the $k$th state,
defined by the set
$\{U_k^+,U_k^-,V_k^+,V_k^-,\Omega_k\}$, satisfies the HFB equation
(\ref{eq143}), then the $\bar{k}$th state, corresponding to the set defined by
$\{U_k^+,-U_k^-,V_k^+,-V_k^-,-\Omega_k\}$, also satisfies the HFB equation
for the same quasiparticle energy $E_k$. Moreover, all wave functions
in cylindrical coordinates are  real.
Contributions of time-reversal states $k$ and $\bar{k}$ are identical
(we assume that the set of occupied states is invariant with respect to the
time-reversal), and we can restrict all summations to positive values
of $\Omega_k$ while multiplying total results by a factor
two. In a similar way, one can see that due to the assumed reflection
symmetry, only positive values of $z$ need to be considered.

\subsubsection{HO and THO Wave Functions}
\label{sec25}

The solution of the HFB equation (\ref{eq143}) is obtained by expanding
the quasiparticle function (\ref{hfbs}) in a given complete set of
basis wave functions that conserve   axial symmetry and parity.
 The
program HFBTHO \codeversion{} is able to do so for the two basis sets of
wave functions: HO and THO.

The HO set consists of eigenfunctions of a single-particle
Hamiltonian for an axially deformed harmonic oscillator potential.
By using the standard oscillator constants:
\begin{equation}
\beta_z=\frac{1}{b_z}=\left(\frac{m\omega_z}{\hbar}\right)^{1/2}, ~~~
\beta_\bot=\frac{1}{b_\bot}=\left(\frac{m\omega_\bot}{\hbar}\right)^{1/2},
\label{hob}
\end{equation}
and auxiliary variables
\begin{equation}
\xi=z \beta_z,~~~\eta=r^2 \beta_\bot^2,
\label{hocoor}
\end{equation}
the HO eigenfunctions are written explicitly as
\begin{equation}
\Phi_\alpha({\bf r},\sigma)=\psi^\Lambda_{n_r}(r) \psi_{n_z}(z)
\frac{e^{\imath\Lambda\varphi}}{\sqrt{2\pi}}\chi_\Sigma(\sigma),
\label{howf}
\end{equation}
where
\begin{equation}
\begin{array}{rcl}
\psi^\Lambda_{n_r}(r) &=&\beta_\bot \tilde{\psi}^\Lambda_{n_r}(\eta) =
N^\Lambda_{n_r}\beta_\bot\sqrt{2}\eta^{|\Lambda|/2}
e^{-\eta/2}L^{|\Lambda|}_{n_r}(\eta),\\ && \\
\psi_{n_z}(z) &=&\beta_z^{1/2} \tilde{\psi}_{n_z}(\xi) = N_{n_z}\beta_z^{1/2}e^{-\xi^2/2}
H_{n_z}(\xi).
\label{hopol}
\end{array}
\end{equation}
$H_{n_z}(\xi)$ and $L^\Lambda_{n_r}(\eta)$ denote the Hermite and
associated Laguerre polynomials \cite{[Abr70]}, respectively,
and the normalization factors read
\begin{equation}
N_{n_z}=\left(\frac{1}{\sqrt{\pi}2^{n_z}n_z!}\right)^{1/2}
~~~~\mbox{and}~~~~
N^\Lambda_{n_r}=\left(\frac{n_r!}{(n_r+|\Lambda|)!}\right)^{1/2}.
\label{honorm}
\end{equation}
The set of quantum numbers $\alpha=\{n_r,n_z,\Lambda,\Sigma\}$
includes the numbers of nodes, $n_r$ and $n_z$, in the $r$ and $z$
directions, respectively, and the projections on the $z$ axis,
$\Lambda$ and $\Sigma$, of the angular momentum operator and the spin.

The HO energy associated with the HO state (\ref{howf}) reads
\begin{equation}
\epsilon_\alpha=(2n_r+|\Lambda|+1)\hbar\omega_\bot+(n_z+\tfrac{1}{2})\hbar\omega_z,
\label{hoe}
\end{equation}
and the basis used by the code consists of
$M_0$=$(N_{sh}$+$1)(N_{sh}$+$2)(N_{sh}$+$3)/6$ states having the lowest
energies $\epsilon_\alpha$ for the given frequencies
$\hbar\omega_\bot$ and $\hbar\omega_z$. In this way, for the
spherical basis, i.e., for $\hbar\omega_\bot$=$\hbar\omega_z$, all HO shells
with the numbers of quanta $N$=0\ldots$N_{sh}$ are included in the basis. When
the basis becomes deformed, $\hbar\omega_\bot$$\neq$$\hbar\omega_z$,
the code selects the lowest-HO-energy basis states by checking the HO
energies of all states up to 50 HO quanta. Note that in this case the
maximum value of the quantum number $\Omega_k$, and the number of
blocks in which the HFB equation is diagonalized, see
Sec.~\ref{sec26}, depend on the deformation of the basis.

The THO set of basis wave functions consists of transformed harmonic
oscillator functions, which are generated by applying the local
scale transformation (LST) \cite{[Sto83],[Sto88a],[Sto91]} to the HO
single-particle wave functions (\ref{howf}). In the axially deformed
case, the LST acts only on the cylindrical coordinates $r$ and $z$, i.e.,
\begin{equation}
\begin{array}{llll}
r  & \longrightarrow  & r^{\prime }\equiv r^{\prime }(r ,z) & =%
r \,\frac{f({\cal R})}{{\cal R}}, \\[1ex]
z & \longrightarrow  & z^{\prime }\equiv z^{\prime }(r ,z) & =
z \,\frac{f({\cal R})}{{\cal R}},
\end{array}
\label{clst}
\end{equation}
and the resulting THO wave functions read
\begin{equation}
\Phi_\alpha({\bf r},\sigma)=\sqrt{\frac{f^{2}({\cal R})}{
{\cal R}^{2}}\frac{\partial f({\cal R})}{\partial {\cal R}}} \psi^\Lambda_{n_r}\left( \frac{r}{{\cal
R}}f({\cal R})\right) \psi_{n_z}\left( \frac{z}{{\cal R}}f({\cal
R})\right)
\frac{e^{\imath\Lambda\varphi}}{\sqrt{2\pi}}\chi_\Sigma(\sigma),
\end{equation}
where
\begin{equation}
{\cal R}=\sqrt{\frac{z^{2}}{b_{z}^{2}}+\frac{r^{2}}{b_\bot^{2}}}, \label{hoass3}
\end{equation}
and $f({\cal R})$ is a scalar LST function. In the code HFBTHO
\codeversion, function $f({\cal R})$ is chosen as in
Ref.~\cite{[Sto03]}. It transforms the incorrect Gaussian
asymptotic behavior of deformed HO wave functions into the correct
exponential form. Below, we keep the same notation $\Phi_\alpha({\bf
r},\sigma)$ for both HO and THO wave functions, because expressions
in which they enter are almost identical in both cases and are valid
for both HO and THO variants.

\subsection{HFB Diagonalization in Configurational Space}
\label{sec26}

We use the same basis wave functions to expand upper and lower components
of the quasiparticle states, i.e.,
\begin{equation}
\begin{array}{rcl}
U_k({\bf r},\sigma,\tau)&=&\displaystyle\chi_{q_k}(\tau)\sum_\alpha U_{k\alpha}\Phi_\alpha({\bf r},\sigma),\\&&\\
V_k({\bf r},\sigma,\tau)&=&\displaystyle\chi_{q_k}(\tau)\sum_\alpha V_{k\alpha}\Phi_\alpha({\bf r},\sigma),\\
\end{array}
\label{wfexp}
\end{equation}
where  $\Phi_\alpha({\bf
r},\sigma)$ are the HO or THO basis states. Note
that the same basis $\Phi_\alpha({\bf r},\sigma)$ is used for protons and
neutrons.

Inserting expression (\ref{wfexp}) into the HFB equation
(\ref{eq143}) and using the orthogonality of the basis states, we find
that the expansion coefficients have to be eigenvectors of the HFB
Hamiltonian matrix
\begin{equation}
\left(
\begin{array}{cc}
h^{(q_k)}-\lambda^{(q_k)} & \tilde{h}^{(q_k)} \\
\tilde{h}^{(q_k)} & -h^{(q_k)}+\lambda^{(q_k)}
\end{array}
\right) \left(
\begin{array}{l}
U_{k} \\
V_{k}
\end{array}
\right) =E_{k}\left(
\begin{array}{l}
U_{k} \\
V_{k}
\end{array}
\right) \;,
\label{hfbmeq}
\end{equation}
where the quasiparticle energies $E_{k}$, the
chemical potential $\lambda^{(q_k)} $, and the matrices
\begin{equation}
h^{(q)}_{\alpha\beta}=\langle\Phi_\alpha|h_q|\Phi_\beta\rangle
~~~~\mbox{and}~~~~
\tilde{h}^{(q)}_{\alpha\beta}=\langle\Phi_\alpha|\tilde{h}_q|\Phi_\beta\rangle
\label{hfbm}
\end{equation}
are defined for a given proton ($q_k$=$+1/2$) or neutron ($q_k$=$-1/2$)
block.

Proton and neutron blocks are decoupled and can be diagonalized separately.
Furthermore, in the case of
axially deformed nuclei considered here,
$\Omega_k$=$\Lambda_k$+$\Sigma_k$ is a good quantum number and,
therefore, matrices $h^{(q)}_{\alpha\beta}$ and
$\tilde{h}^{(q)}_{\alpha\beta}$ are block diagonal, each block being
characterized by a given value of $\Omega$. Moreover, for the case
of conserved parity considered here, $\pi$=$(-1)^{n_z+\Lambda}$ is
also a good quantum number, and each of the  $\Omega_k$ blocks falls
into two sub-blocks characterized by the values of $\pi$=$\pm1$.
Finally, due to the time-reversal symmetry, the Hamiltonian
matrices need to be constructed for positive values of $\Omega_k$
only.

\subsection{Calculations of Matrix Elements}
\label{sec27}

As discussed in Sect.~\ref{sec22}, local densities (\ref{locden}) and
average fields, (\ref{hppq}) and (\ref{mt}), are calculated in the
coordinate space. Therefore, calculation of matrix elements
(\ref{hfbm}) amounts to calculating appropriate spatial integrals in
the cylindrical coordinates $r$ and $z$. In practice, the integration
is carried out by using the Gauss quadratures \cite{[Abr70]} for 22
Gauss-Hermite points in the $z>0$ direction and 22 Gauss-Laguerre points
in the $r$ direction. This gives a sufficient
accuracy for calculations up to $N_{sh}=40$.

In the case of the HO basis functions, the integration is performed
by using the Gauss integration points, $\xi_n$ and $\eta_m$, for which
the local densities and fields have to be calculated at the
mesh points of $z_n$=$b_z\xi_n$ and $r_m$=$b_\bot\eta_m^{1/2}$. As an
example, consider the following diagonal matrix element of the
potential $U_q(r,z)$ (\ref{mt}),
\begin{equation}
U^q_{\alpha\alpha}=\int\limits_{-\infty}^{\infty}
dz\int\limits_{0}^{\infty} r d r~ U_q(z,r)~\psi_{n_{z}}^2(z) {\psi_{n_r}^\Lambda}^2(r).
\label{aaa}
\end{equation}
Inserting here the HO functions $\psi_{n_{z}}^2(z)$ and
${\psi_{n_r}^\Lambda}^2(r)$ (\ref{howf}), and changing the integration
variables to dimensionless variables $\xi$ and $\eta$, the above
matrix element reads
\begin{equation}
U^q_{\alpha\alpha}=\int\limits_{-\infty}
^{\infty}d\xi\int\limits_{0}^{\infty} d\eta\, \tilde{U}_q(\xi,\eta)
\tilde{\psi}_{n_z}^2 (\xi) \tilde{\psi}_{n_r}^{\Lambda~2}(\eta),
\label{aaa1}
\end{equation}
where
\begin{equation}
\tilde{U}_q(\xi,\eta)=\tfrac{1}{2}
                       U_q(\xi b_z,\sqrt{\eta}\, b_\bot).
\end{equation}
Here, the Gauss integration quadratures can be directly applied, because
the HO wave functions contain appropriate exponential profile functions.

The situation is a little bit more complicated in the case of the THO basis
states where, before calculating, one has to change variables with
respect to the LST functions $f({\cal R})$. For example, let us
consider the same matrix elements (\ref{aaa}) but in THO representation:
\begin{equation}
U^q_{\alpha\alpha}= \displaystyle
\int\limits_{-\infty}^{\infty}dz\int\limits_{0}^{\infty}r d r~ U_q(z,r)
\left[  \frac{f^{2}(\mathcal{R})}{\mathcal{R}^{2}}\frac
{df(\mathcal{R})}{d\mathcal{R}}\right]  \psi_{n_{z}}^{2}\left(
\frac{zf(\mathcal{R})}{\mathcal{R}}\right)
{\psi_{n_{r}}^\Lambda}^2\left(  \frac{r^{2}f^{2}(\mathcal{R})}{\mathcal{R}^{2}}\right)  .
\end{equation}
Introducing new dimensionless variables
\begin{equation}
\displaystyle
\xi=\frac{z}{b_{z}}\frac{f(\mathcal{R})}{\mathcal{R}},~~~
\eta=\frac{r^{2}}{b_{\bot}^{2}}\frac{f^{2}(\mathcal{R})}{\mathcal{R}^{2}},
\end{equation}
for which we have
\begin{equation}
d\xi~d\eta=\frac{2}{b_{z}b_{\bot}^{2}}\left[  \frac{f^{2}(\mathcal{R}
)}{\mathcal{R}^{2}}\frac{df(\mathcal{R})}{d\mathcal{R}}\right]  r d r~ dz,
\end{equation}
the matrix elements have the form of integrals, which are exactly identical
to those in the HO basis (\ref{aaa1}), after changing the function $\tilde{U}_q(\xi,\eta)$ to
\begin{equation}
\tilde{U}_q(\xi,\eta)=\tfrac{1}{2}
       U_q\left(\xi b_z\tfrac{\mathcal{R}}{f(\mathcal{R})},
        \sqrt{\eta}\, b_\bot\tfrac{\mathcal{R}}{f(\mathcal{R})}\right).
\end{equation}
The calculation of matrix elements corresponding to derivative terms in
the Hamiltonian (\ref{hppq}) can be performed in an analogous way,
after the  derivatives of the Jacobian,
$\tfrac{f^{2}(\mathcal{R}
)}{\mathcal{R}^{2}}\tfrac{df(\mathcal{R})}{d\mathcal{R}}$, are taken
into account.

\subsection{Calculation of Local Densities}
\label{sec27a}

After diagonalizing the HFB equation (\ref{hfbmeq}),
local densities are calculated as
\begin{equation}
\begin{array}{lll}
\rho ({\bf r}\sigma,{\bf r}^{\prime }\sigma^{\prime }) & = & \sum\limits_{\alpha \beta
}\rho_{\alpha \beta }
~\Phi_{\alpha }^{\ast }({\bf r}\sigma)\Phi_{\beta }({\bf r}^{\prime }\sigma^{\prime })\;, \\
\tilde{\rho} ({\bf r}\sigma,{\bf r}^{\prime }\sigma^{\prime }) & = & \sum\limits_{\alpha \beta
}\tilde{\rho}_{\alpha \beta }
~\Phi_{\alpha }^{\ast }({\bf r}\sigma)\Phi_{\beta }({\bf r}^{\prime }\sigma^{\prime })\;,
\end{array}
\label{rhokconfig}
\end{equation}
where $\Phi_\alpha({\bf r}\sigma)$ denotes the HO or THO basis wave functions,
and the matrix elements of mean-field and pairing density matrices read
\begin{equation}
\rho _{\alpha \beta }=  \sum\limits_{k}V_{\alpha
k}^{\ast }V_{\beta k}, ~~~
\tilde{\rho}_{\alpha \beta }= - \sum\limits_{k}
V_{\alpha k}^{\ast }U_{\beta k}\;.
\label{rhokconfig1}
\end{equation}

The HFB calculations for zero-range pairing interaction give
divergent energies when increasing the number of quasiparticle states
in the sums of Eq.~(\ref{rhokconfig1}) (see discussion in
Ref.~\cite{[Dob96]}). Therefore, they invariably require a truncation
of quasiparticle basis by defining a cut-off quasiparticle energy
and including all quasiparticle states only up to
this value.

The choice of an appropriate cut-off procedure has been discussed
in \cite{[Dob84]}. After each iteration, performed with a given
Fermi energy $\lambda$, one calculates an equivalent spectrum
$\bar{e}_{k}$ and pairing gaps $\bar{\Delta}_{k}$:
\begin{equation}
\begin{tabular}{l}
$\bar{e}_{k}=(1-2N_{k})E_{k},$ \\
\  \\
$\bar{\Delta}_{k}=2E_{k}\sqrt{N_{k}(1-N_{k})},$%
\end{tabular}
\label{ENnbar}
\end{equation}
where $N_{k}$ denotes the norm (\ref{Ntot}) of the lower HFB wave
function. Using this spectrum and pairing gaps, the Fermi energy is
readjusted to obtain the correct value of particle number, and this
new value is used in the next HFB iteration.

Due to the similarity between the equivalent spectrum
$\bar{e}_{k}$ and the single-particle energies, one can take
into account only those quasiparticle states for which
\begin{equation}
\bar{e}_{k}\leq \bar{e}_{\max },  \label{cutoff}
\end{equation}
where $\bar{e}_{\max }$$>$0 is a parameter defining the amount of
the positive-energy phase space taken into account. Since all
hole-like quasiparticle states, $N_{k}$$<$1/2, have negative values of $\bar{%
e}_{k}$, condition (\ref{cutoff}) guarantees that they are all
taken into account. In this way, a global cut-off prescription is
defined which fulfills the requirement of taking into account the
positive-energy phase space as well as all quasiparticle states
up to the highest hole-like quasiparticle energy. In the code, a
default value of $\bar{e}_{\max }$=60 MeV is used.

\subsection{Coulomb Interaction}
\label{sec28}

In the case of proton states, one has to add to the central potential
the direct Coulomb field
\begin{equation}
V^C_d({\bf r})=e^2 \int d^3 {\bf r^\prime}
\frac{\rho_p({\bf r^\prime})}{|{\bf r}-{\bf r^\prime}|},
\label{could}
\end{equation}
as well as the exchange Coulomb field, which in the present implementation
is treated  within  the Slater approximation:
\begin{equation}
V^C_{ex}({\bf r})=  -e^2\left(\tfrac{3}{\pi}\right)^{1/3}
                  \rho_p^{1/3}({\bf r}) .
\end{equation}

The integrand in the direct term (\ref{could}) has a logarithmic
singularity at the point ${\bf r}$=${\bf r^\prime}$. A way to bypass
this difficulty is to use the Vautherin prescription
\cite{[Vau73]}, i.e., to employ the identity
\begin{equation}
\triangle_{{\bf r}^\prime}|{\bf r}-{\bf r^\prime}|=
2/|{\bf r}-{\bf r^\prime}|,
\label{vau}
\end{equation}
and then integrate by parts the integral in Eq.~(\ref{could}). As a
results, one obtains a singularity-free expression
\begin{equation}
V^C_d({\bf r})=\frac{e^2}{2} \int d^3 {\bf r^\prime}~
|{\bf r}-{\bf r^\prime}|~\triangle_{{\bf r}^\prime} \rho_p({\bf r^\prime}).
\label{vau1}
\end{equation}
In cylindrical coordinates, after integrating over the azimuthal
angle $\varphi$, one finds
\begin{equation}\displaystyle
V^C_d(r^\prime,z^\prime)=2 e^2 \int_0^\infty r d r \int_{-\infty}^\infty d z~
\sqrt{d(r,z)}~
E\left(\frac{4rr^\prime}{d(r,z)}\right)
~\triangle \rho_p(r,z),
\label{vau2}
\end{equation}
where $d(r,z)=\left[(z-z^\prime)^2+(r+r^\prime)^2\right]$ and $E(x)$
is the complete elliptic integral of the second kind that can be
approximated by a standard polynomial formula \cite{[Abr70]}.

Equivalently, one can use the prescription developed originally
for calculations with the finite-range (Gogny) force \cite{[Dob96]}.
It consists
of expressing the Coulomb force as a sum of Gaussians:
\begin{equation}
\displaystyle
\frac{ 1 }{ \vert {\bf r} - {\bf r}' \vert } =
\frac{ 2 }{ \sqrt{\pi}} \int_{0}^{\infty} \frac{\text{d} \mu }{ \mu^{2}}~
e^{-\frac{({\bf r} - {\bf r}')^2 }{ \mu^2}},
\label{go}
\end{equation}
which gives
\begin{equation}
\displaystyle
V^C_d({\bf r})=e^2
\frac{ 2 }{ \sqrt{\pi}} \int_{0}^{\infty} \frac{\text{d} \mu }{ \mu^{2}}~
I_{\mu} ({\bf r}),
\label{go1}
\end{equation}
where the integral
\begin{equation}\displaystyle
I_{\mu} ({\bf r})  =  \int\text{d}^{3} {\bf r}'
e^{-\frac{({\bf r} - {\bf r}')^2 }{ \mu^2}}
\; \rho ({\bf r}')
      \label{go2}
\end{equation}
can be easily calculated in cylindrical coordinates. After
integrating over the azimuthal angle $\varphi$, one finds
\begin{equation}\displaystyle
I_{\mu}(r^\prime,z^\prime)= 2 \pi \int_0^\infty r d r \int_{-\infty}^\infty
d z~
  e^{-\frac{r^2 + r^{\prime 2} + {\left( z - z^\prime \right) }^2}{\mu^2}}
I_0\left(\frac{2 r r^\prime}{\mu^2}\right)~
  \rho_p(r,z),
\label{go3}
\end{equation}
where $I_0(x)$ is the Bessel function that can also be approximated
by a standard polynomial formula \cite{[Abr70]}.

In order to perform the remaining one-dimensional integration in
Eq.~(\ref{go1}), the variable $\mu$ is changed to
\begin{equation}
\xi = b / \sqrt{b^2 + \mu^{2}} ,
\label{go4}
\end{equation}
where $b$ is the largest of the two HO lengths $b_z$
and $b_\bot$.  This change of variable is very convenient, since then
the range of integration becomes [0, 1]. The integral (\ref{go1}) is
accurately computed by using a 30-point Gauss-Legendre quadrature
with respect to $\xi$.

We have tested the precision of both prescriptions, Eqs.~(\ref{vau1}) and
(\ref{go1}), and checked that the second one gives better results
within the adopted numbers of Gauss-Hermite and Gauss-Laguerre points
that are used for calculating proton densities. Therefore, in the
code HFBTHO {\codeversion} this second prescription is used,
while the first one remains in the code, but is inactive.

\subsection{Lipkin-Nogami Method}
\label{sec29}

The LN method constitutes an efficient method for approximately
restoring the particle numbers before variation \cite{[Lip60]}. With
only a slight modification of the HFB procedure outlined above, it is
possible to obtain a very good approximation for the optimal HFB
state, on which exact particle number projection then has to be
performed \cite{[Dob93],[Egi02]}.

In more detail, the LN method is implemented by performing the HFB
calculations with an additional term included in the HF
Hamiltonian,
\begin{equation}
\label{hprime}
h' = h - 2\lambda_2(1-2\rho),
\end{equation}
and by iteratively calculating the parameter $\lambda_2$
(separately for neutrons and protons) so as to properly describe
the curvature of the total energy as a function of particle number.
For an arbitrary two-body interaction $\hat{V}$, $\lambda_2$  can
be calculated from the particle-number dispersion according to
\cite{[Lip60]},
\begin{equation}
\lambda_{2}=\frac {\langle 0| \hat{V} |4\rangle\langle 4| \hat
N^{2} |0\rangle}{\langle0|\hat N^{2}|4 \rangle\langle4|\hat N^{2}
|0\rangle} ~,
\end{equation}
where $|0\rangle$ is the quasiparticle vacuum, $\hat{N}$ is the
particle number operator,  and $|4\rangle\langle4|$ is the
projection operator onto the 4-quasiparticle space. On evaluating
all required matrix elements, one obtains \cite{[Flo97]}
\begin{equation}\label{ll2}
\lambda_{2}=\frac {4{\rm Tr} \Gamma^{\prime} \rho(1-\rho) + 4{\rm
Tr}\Delta^{\prime} (1-\rho)\kappa} {8\left[{\rm Tr}\rho (1-\rho
)\right]^{2}-16{\rm Tr}\rho^{2}(1-\rho)^{2}} ~,
\end{equation}
where the potentials
\begin{equation}\label{eq42}
\begin{array}{rcl}
\Gamma^{\prime}_{\alpha \alpha^{\prime}} &=& \sum_{\beta \beta^{\prime}}V_{\alpha \beta
\alpha^{\prime} \beta^{\prime}}(\rho(1-\rho))_{\beta^{\prime} \beta}, \\ ~ && \\
\Delta^{\prime}_{\alpha \beta} &=&\thalf\sum_{\alpha^{\prime}
\beta^{\prime}}V_{\alpha \beta \alpha^{\prime} \beta^{\prime}}(\rho
\kappa)_{\alpha^{\prime} \beta^{\prime}},
\end{array}
\end{equation}
can be calculated in
a full analogy to $\Gamma$ and $\Delta$ by replacing  $\rho$ and
$\kappa$  by $\rho(1-\rho)$ and
$\rho\kappa$, respectively. In the case of the seniority-pairing interaction
with strength $G$,
Eq.~(\ref{ll2})  simplifies to
\begin{equation}\label{ll3}
\lambda_{2}=\frac{G}{4} \frac {{\rm Tr} (1-\rho)\kappa~ {\rm Tr} \rho \kappa  - 2~{\rm Tr} (1-\rho)^2
\rho^2} {\left[{\rm Tr}\rho (1-\rho )\right]^{2}-2~{\rm Tr}\rho^{2}(1-\rho)^{2}}.
\end{equation}

An explicit calculation of $\lambda_{2}$ from Eq.~(\ref{ll2}) requires
calculating new sets of fields (\ref{eq42}),
which is rather cumbersome. However, we have found \cite{[Sto04a]}
that Eq.~(\ref{ll2}) can be well approximated by the seniority-pairing
expression (\ref{ll3})
with the effective strength
\begin{equation}
G=G_{\text{eff}} = -\frac{\bar{\Delta}^2}{E_{\text{pair}}}\,
\end{equation}
determined from the pairing energy
\begin{equation}
E_{\text{pair}} = -\tfrac{1}{2}{\rm Tr}\Delta \kappa \,
\end{equation}
and the average pairing gap
\begin{equation}
\bar{\Delta}  = \frac{{\rm Tr}\Delta \rho}{{\rm Tr}\rho} \, .
\end{equation}
Such a procedure is implemented in the code HFBTHO \codeversion.

\subsection{Particle-Number Projection After Variation }
\label{sec210}

Introducing the particle-number projection operator for $N$ particles,
\begin{equation}
P^{N}=\tfrac{1}{2\pi }\int d\phi \ e^{i\phi (\hat{N}-N)},  \label{E22}
\end{equation}
where $\hat{N}$ is the number operator, the average HFB energy of the
particle-number projected state can be expressed as an integral over the
gauge angle $\phi$ of the Hamiltonian matrix elements between states
with different gauge angles \cite{[She00a],[She02]}. In particular, for the Skyrme-HFB
method implemented here, the particle-number projected
 energy can be written as \cite{[Sto04b],[Sto04a]}
\begin{equation}
\textsf{E}^{N}[\rho,\tilde{\rho}]=\frac{\left\langle \Phi |HP^{N}|\Phi \right\rangle }{
\left\langle \Phi |P^{N}|\Phi \right\rangle }
=\int d\phi ~y(\phi )\int d^3{\bf r}~{\cal  H}({\bf r},\phi )~, \label{shfbN}
\end{equation}
where the gauge-angle dependent energy density ${\cal  H}({\bf r},\phi )$
is derived from the unprojected energy density ${\cal  H}({\bf r})$
(\ref{enden}) by simply substituting the
particle and pairing local densities $\rho ({\bf r})$, $\tilde{\rho}({\bf
r})$, $\tau ({\bf r})$, and ${\bf
J}_{ij}({\bf r})$ by their gauge-angle
dependent counterparts $\rho ({\bf r},\phi )$, $\tilde{\rho}({\bf
r},\phi )$, $\tau ( {\bf r},\phi )$, and ${\bf J}_{ij}({\bf r},\phi )$, respectively.
The latter densities are calculated from the gauge-angle dependent density matrices as
\begin{equation}
\begin{array}{c}
\displaystyle{\rho ({\bf r}\sigma ,{\bf r^{\prime }}\sigma^{\prime },\phi )
=\sum_{\alpha\alpha^{\prime }}\rho_{\alpha\alpha^{\prime }}(\phi )~\Phi_{\alpha^{\prime
}}^{\ast }({\bf r^{\prime }},\sigma^{\prime })\Phi_{\alpha}({\bf
r},\sigma )},
\\~~\\
\displaystyle{\tilde{\rho}({\bf r}\sigma ,{\bf r^{\prime }}\sigma^{\prime },\phi
) =\sum_{\alpha\alpha^{\prime }}\tilde{\rho}_{\alpha\alpha^{\prime }}(\phi )~\Phi
_{\alpha^{\prime }}^{\ast }({\bf r^{\prime }},\sigma^{\prime })\Phi
_{\alpha}({\bf r},\sigma )},
\end{array}
\label{penmcp}
\end{equation}
where the gauge-angle dependent matrix elements read
\begin{equation}
\begin{array}{c}
\displaystyle{\rho_{\alpha^{\prime }\alpha}(\phi )
= \sum_\beta C_{\alpha\beta}(\phi )\rho_{\beta \alpha^{\prime }}} ,
\\~~\\
\displaystyle{\tilde{\rho}_{\alpha^{\prime }\alpha}(\phi ) = e^{-i\phi }\sum_\beta C_{\alpha \beta}(\phi )\tilde{\rho}_{\beta\alpha^{\prime }}} ,
\end{array}
\label{gen}
\end{equation}
and depend on the unprojected matrix elements (\ref{rhokconfig1}) and
on the  gauge-angle dependent matrix
\begin{equation}
C(\phi ) =e^{2i\phi }\left[ 1+\rho (e^{2i\phi }-1)\right]^{-1}.
\label{cc}
\end{equation}
Function $y(\phi )$ appearing in Eq. (\ref{shfbN}) is defined as
\begin{equation}
y(\phi ) =\frac{x(\phi )}{\int d\phi^{\prime }\,x(\phi^{\prime })}
~~~~\mbox{for}~~~~
x(\phi )=\tfrac{1}{2\pi }\frac{e^{-i\phi N}\det (e^{i\phi}I)}{\sqrt{\det C(\phi )}},
\label{ynm}
\end{equation}
where $I$ is the unit matrix.

Since the gauge-angle dependent matrices (\ref{penmcp}) and
(\ref{gen}) are all diagonal in the same canonical basis that
diagonalizes the unprojected density matrices (\ref{rhokconfig1}),
all calculations are very much simplified when they are performed in
the canonical basis. In particular, in the canonical basis the
matrices (\ref{gen}) read
\begin{equation}
\rho_{\mu}(\phi) =\displaystyle \frac{e^{2\imath \phi}v_{\mu}^{2}}{u_{\mu
}^{2}+e^{2\imath \phi }v_{\mu}^{2}}
~~~~~\mbox{and}~~~~
\tilde{\rho}_{\mu}(\phi) =\displaystyle \frac{e^{\imath \phi}
u_{\mu}v_{\mu}}{u_{\mu}^{2}+e^{2\imath \phi}v_{\mu}^{2}},
\end{equation}
while the function $x(\phi)$ can be calculated as
\begin{equation}
x(\phi) =\displaystyle{e^{-\imath
N\phi}\prod_{\mu >0}\left( u_{\mu}^{2}+e^{2\imath \phi}v_{\mu}^{2}\right) },
  \label{E402}
\end{equation}
where $v_{\mu}$ and $u_{\mu}$ ($v_{\mu}^2+u_{\mu}^2=1$) are the usual
 canonical basis occupation amplitudes.

All the above expressions apply to independently restoring the proton and neutron numbers,
so, in practice, integrations over two gauge angles have to be
simultaneously implemented. In practice, these integrations are
carried out by using a simple discretization method, which amounts
to approximating the projection operator (\ref{E22}) by a
double sum \cite{[Fom70]}, i.e.,
\begin{equation}
P^{NZ}=\frac{1}{L}\sum_{l_{n}=0}^{L-1}\ e^{i\phi
_{n}(\hat{N}-N)}\frac{1}{L} \sum_{l_{p}=0}^{L-1}\ e^{i\phi
_{p}(\hat{Z}-Z)},  \label{E22NZS}
\end{equation}
where
\begin{equation}
\phi_{q}=\frac{\pi }{L}l_{q},~~~~q=n,p.  \label{E22FI}
\end{equation}
Usually no more than $L=9$ points are required for a precise
particle number restoration.

\subsection{Constraints}
\label{sec228}

In the code HFBTHO {\codeversion}, the HFB energy (\ref{shfb}) can be minimized
 under the  constraint of a fixed quadrupole moment. This option
 should be used  if one is interested in the potential energy surface
 of a nucleus along the quadrupole collective coordinate.
The quadrupole  constraint is assumed in the standard
quadratic form \cite{[Flo73]}:
  \begin{equation}\label{conn}
   {E}^{Q} = C_{Q}
                \left(\langle\hat Q\rangle
                           - \bar Q\right)^2,
 \end{equation}
where $\langle\hat Q\rangle$ is the average value
of the mass-quadrupole-moment operator,
\begin{equation}\label{QQQ}
\hat{Q}=2z^2-r^2,
\end{equation}
$\bar Q$ is the constraint value of the quadrupole
moment, and $C_{Q}$ is the stiffness constant.

\section{Program HFBTHO \codeversion}
\label{sec3}

The code HFBTHO \codeversion{} is written in Fortran~95 with {\tt MODULE}
definitions that specify all common arrays and variables for other
subroutines by using the {\tt USE} statements. Integer and real
types of variables are automatically detected for the particular computer
through the {\tt KIND } statements. The code is entirely portable. It
contains all initial data and no references to
external subroutines or libraries are made.

The code requires one input data file ({\tt tho.dat}). Optionally, in
case one wants to restart calculations from a previous run, two more
files, {\tt dnnn\_zzz.hel} and/or {\tt dnnn\_zzz.tel}, are required
as described below. Also optionally, if one wants to run the code for
user-defined Skyrme-force parameters, file {\tt forces.dat} is required.

The results are printed on the standard output and also recorded in
the file {\tt thoout.dat}. The main results are also recorded in the
files {\tt hodef.dat} (HO basis) and {\tt thodef.dat} (THO basis),
where one line is written for every nucleus calculated, producing a
concise table of results suitable for further analyses. Files {\tt
hodef.dat} and {\tt thodef.dat} are also used when restarting the
given calculation after an abnormal termination (CPU time limit or
system crash). Namely, before performing a given run, the code always
checks if the line corresponding to this run is present or not in the
file {\tt hodef.dat} or {\tt thodef.dat}. If this is the case, the code does not
repeat the calculation for the given run, and only the runs which have
not been completed are executed. Due to this implementation, if the
user wishes to rerun the same input data file, files {\tt hodef.dat}
and {\tt thodef.dat} have to be first removed from the current
directory.

\subsection{General Structure of the Code}
\label{sec31}

The code runs, in sequence, the set of main subroutines listed
in Table \ref{tab1}. If multiple
runs are requested in a single input data file, the code always repeats
the whole sequence of calls from the beginning to end, including an
initialization of all variables and data.

\begin{table}
\caption[T1]{List of main subroutines constituting the code HFBTHO \codeversion}
\label{tab1}
\vspace*{0.5cm}
\begin{tabular}{|l|p{12.5cm}|}
\hline
Subroutine & Task \\
\hline
{\tt DEFAULT}& Initializes all variables (initially, or after the previous run). \\
{\tt READ INPUT}& Reads parameters from the input data file {\tt tho.dat}.\\
{\tt PREPARER}& Initializes variables according to the user's request defined
in the input data file.\\
{\tt BASE0}& Determines the HO configurational space and dimensions of allocatable arrays.\\
{\tt THOALLOC}& Allocates memory required for the given run of the code.\\
{\tt BASE}& Calculates and stores properties of the configurational space and all associated quantum numbers.\\
{\tt GAUPOL}& Calculates and stores the HO basis wave functions.\\
{\tt INOUT}& Sets or reads (optional) initial densities, fields, and matrix elements.\\
{\tt ITER}& Main iteration loop for the HFB+HO calculation, which is repeated until convergence is met. It includes the following subroutines: \\[2ex]
& \begin{tabular}{@{}lp{9.7cm}}
{\tt DENSIT}& Calculates densities in coordinate space.\\
{\tt FIELD}& Calculates mean fields in coordinate space.\\
{\tt GAMDEL}& Calculates the particle-hole  and pairing Hamiltonian matrices.\\
{\tt EXPECT}& Calculates average values of observables.\\
{\tt HFBDIAG}& Diagonalizes the HFB equation.\\[2ex]
\end{tabular} \\
{\tt F01234}& After the HFB+HO solution is found, calculates
     the THO basis wave functions, which replace the HO ones.\\
{\tt ITER}& Main iteration loop for the HFB+THO calculation, which is repeated until convergence is met. The same subroutine and sequence of calls is used as above.\\
{\tt RESU}& Calculates all required physical characteristics and canonical basis properties, and performs the particle number projection.\\
{\tt INOUT}& Records the final densities, fields, and matrix elements for feature use (optional).\\
\hline
\end{tabular}
\end{table}

\subsection{Input Data File}
\label{sec32}

Input data are read from file {\tt tho.dat}, which is shown in
Table~\ref{table1}. The file consists of the
first line, which contains only two numbers, below referred to as I1
and I2, followed by a sequence of identical lines, each of them
defining one specific run of the code. All numbers containing a dot
are type real, and those without a dot are type integer. In
quotations there are four-character strings giving acronyms of the
Skyrme forces. The code uses free format, so at least one space is
needed in order to separate the input numbers.

\begin{table}
\centering
\caption{Input data file {\tt tho.dat}.}
\label{table1}
\vspace*{0.5cm}
\begin{tabular}{|r@{\;\;}r@{\;\;}r@{\;\;}r@{\;\;}r@{\;\;}r@{\;\;}r@{\;\;}r@{\;\;}r@{\;\;}r@{\;\;}r@{\;\;}r@{\;\;}r@{\;\;}r@{\;\;}r@{\;\;}r@{\;\;}r@{\;\;}r@{\;\;}r@{\;\;}|}
\hline
   (a)&(b)&(c)&(d)&(e)&(f)&(g)&(h)&(i)~~~~&(j)&(k)&(l)&(m)&(n)&(o)&(p)&(q)&(r)&(s)  \\
  \hline
$-$20 & 20 &  &  &  &  &  &  &  &  &  &  &  &  &  &  &  &  &  \\
   20 & $-$2. & 0. &$-$1 & 300 & 1 & 70 & 50 & 'SLY4' &$-$1 & 1 & 0 & 0.26 & 0.5 & 9 & 0 &   2 &   2&0.0001  \\
   20 & $-$2. & 0. &$-$1 & 300 & 1 & 72 & 50 & 'SLY4' &$-$1 & 1 & 0 & 0.26 & 0.5 & 9 & 0 &   2 &$-$2&0.0001  \\
   20 & $-$2. & 0. &$-$1 & 300 & 1 & 74 & 50 & 'SLY4' &$-$1 & 1 & 0 & 0.26 & 0.5 & 9 & 0 &$-$2 &   2&0.0001  \\
   20 & $-$2. & 0. &$-$1 & 300 & 1 & 76 & 50 & 'SLY4' &$-$1 & 1 & 0 & 0.26 & 0.5 & 9 & 0 &$-$2 &$-$2&0.0001  \\
$-$14 & $-$2. & 0. &$-$1 & 300 & 1 & 78 & 50 & 'SKP ' &$-$1 & 1 & 0 & 0.26 & 0.5 & 9 & 0 &   4 &   4&0.0001  \\
    0 & $-$2. & 0. &$-$1 & 300 & 1 & 70 & 50 & 'SLY4' &$-$1 & 1 & 0 & 0.26 & 0.5 & 9 & 0 &   0 &   0&0.0001  \\
  \hline
 \end{tabular}
\end{table}

The code has three main regimes:

\vspace*{0.5cm}
\begin{tabular}{lll}
(i)   &{\bf nucleus-after-nucleus}, & defined by I1$<$0, \\
(ii)  &{\bf file-after-file},       & defined by I1$=$0, \\
(iii) &{\bf chain-after-chain},     & defined by I1$>$0. \\
\end{tabular}
\vspace*{0.5cm}

In the {\bf nucleus-after-nucleus} regime, the code ignores the
values of $|$I1$|$ and I2, and then performs one run for each line of the
input data file that follows the first line. This is the simplest
and most often used regime, illustrated by the example given in
Table~\ref{tab1}.

In the {\bf file-after-file} regime, the code ignores the value of
I2, and then reads the second line of the input data file, from where
it takes all fields except from the values of {\tt ININ}, $N$, and
$Z$. Then it performs one run for each {\tt dnnn\_zzz} file found
in the current directory. Files {\tt dnnn\_zzz} contain results of
previous runs and are described below.

In the {\bf chain-after-chain} regime, the code reads the second line
of the input data file, from where it takes all fields except from the
values of $N$, and $Z$. Then it performs one run for
each nucleus in the chain of isotones or isotopes located
between the bottom of the stability valley and the drip line.
The bottom of the stability valley is parametrically defined as
\begin{equation}
f(N,Z) \equiv N - Z - 0.006(N+Z)^{5/3} = 0.
\label{stab}
\end{equation}

\begin{itemize}
\item
If I2$>$0, the code calculates the chain of isotopes for the proton
number $Z$=I1, starting with the lowest even neutron number $N$
satisfying $f(N,Z)$$>$0, and then step-by-step increasing the number
of neutrons by two. Calculations continue until the neutron drip line
is reached, and then the program stops.

\item
If I2$<$0, the code calculates the chain of isotones for the neutron
number $N$=I1, starting with the lowest even proton number $Z$
satisfying $f(N,Z)$$<$0, and then step-by-step increasing the number
of protons by two. Calculations continue until the proton drip line
is reached, and then the program stops. \end{itemize}

All lines of the input data file, after the first line, contain 19
fields each. Below we denote these fields by letters {\bf (a)} --
{\bf (s)}, as shown in the header of Table~\ref{table1}. The description
of the fields is as follows:

\begin{itemize}
\item {\bf (a)} Number of oscillator shells $N_{sh}$:
\begin{itemize}
   \item If $N_{sh}>0$, the code prints intermediate results at every iteration.
   \item If $N_{sh}<0$, the code prints results at the first and last iterations only,
                        and the module of the input value is used for $N_{sh}$.
   \item If $N_{sh}=0$, the code stops. This value is used to indicate the end of the input data file.
\end{itemize}
For $N_{sh}>14$, the code always begins with a short, 20-iteration
run using $N_{sh}=14$, and the resulting fields then serve as a
starting point for the calculation with the requested value of $N_{sh}$.
For the THO-basis calculations, use of $N_{sh}<14$ is not
recommended, because precision of the HO density profile can  be
insufficient for a reliable determination of the LST function.
\item {\bf (b)} Oscillator basis parameter $b_0$=$\sqrt{b_z^2+b_\bot^2}$:
\begin{itemize}
   \item If $b_0 >0$, the code uses this given value of $b_0$.
   \item If $b_0 <0$, the code uses the default value of $b_0=\sqrt{2(\hbar^2/2m)/(41 f A^{-1/3})}$ for $f$=1.2.
\end{itemize}
\item {\bf (c)} Deformation $\beta_0$ of the HO basis. The value of $\beta_0$ defines
                the HO oscillator lengths through $b_\bot=b_0q^{-1/6}$, $b_z=b_0q^{1/3}$,
                and $q=\exp \left(3\sqrt{5/(16\pi)} \beta_0\right)$. In particular,
                the value of $\beta_0=0$ corresponds to the spherical HO basis.
\item {\bf (d)} The THO basis control parameter {\tt ILST}:
\begin{itemize}
 \item If {\tt ILST}=    0, the code performs the HO  basis calculation  only. If {\tt ININ}$<0$, the file {\tt dnnn\_zzz.hel} is used as the starting point. If {\tt MAXI}$>$0, at the end of the given run file {\tt dnnn\_zzz.hel} is stored.
 \item If {\tt ILST}= $-$1, the code performs the HO  basis calculation followed by the THO basis calculation. If {\tt ININ}$<0$, the file {\tt dnnn\_zzz.hel} is used as the starting point. If {\tt MAXI}$>$0, at the end of the given run files {\tt dnnn\_zzz.hel} and {\tt dnnn\_zzz.tel} are stored.
 \item If {\tt ILST}=    1, the code performs the THO basis calculation  only. File {\tt dnnn\_zzz.tel} must exist and is used as the starting point (only {\tt ININ}$<$0 is allowed). If {\tt MAXI}$>$0, at the end of the given run file {\tt dnnn\_zzz.tel} is stored.
\end{itemize}
\item {\bf (e)} Maximal number of iterations {\tt MAXI}. If the negative number is read, the absolute value is used.
\begin{itemize}
   \item If {\tt MAXI}$>$0, at the end of the given run files {\tt dnnn\_zzz.hel} and/or {\tt dnnn\_zzz.tel} are stored,
   \item If {\tt MAXI}$<$0, files {\tt dnnn\_zzz.hel} and {\tt dnnn\_zzz.tel} are not stored,
                            and the module of the input value is used for {\tt MAXI}.
\end{itemize}
\item {\bf (f)} The starting-point control parameter {\tt ININ}:
\begin{itemize}
\item If {\tt ININ=}    ~1, the code starts from a default spherical field predefined within the code,
\item If {\tt ININ=}    ~2, the code starts from a default prolate   field predefined within the code,
\item If {\tt ININ=}    ~3, the code starts from a default oblate    field predefined within the code,
\item If {\tt ININ=}  $-$1, the code starts from file {\tt snnn\_zzz.hel} or {\tt snnn\_zzz.tel},
\item If {\tt ININ=}  $-$2, the code starts from file {\tt pnnn\_zzz.hel} or {\tt pnnn\_zzz.tel},
\item If {\tt ININ=}  $-$3, the code starts from file {\tt onnn\_zzz.hel} or {\tt onnn\_zzz.tel}.
\end{itemize}
\item {\bf (g)} Number of neutrons $N$.
\item {\bf (h)} Number of protons  $Z$.
\item {\bf (i)} Skyrme force character*4 acronym, e.g., 'SIII', 'SKP ', 'SLY4', or 'SKM*'. If value 'READ' is read, the code reads the Skyrme force parameters from file {\tt forces.dat}.
                An example of the file {\tt forces.dat} is presented in Table~\ref{tab2}.
\item {\bf (j)} The Lipkin-Nogami control parameter {\tt KINDHFB}:
\begin{itemize}
\item If {\tt KINDHFB}=  ~1, Lipkin-Nogami correction not included,
\item If {\tt KINDHFB}=$-$1, Lipkin-Nogami correction included.
\end{itemize}
\item {\bf (k)} The pairing-force control parameter {\tt IPPFORCE}:
\begin{itemize}
\item If {\tt IPPFORCE}= 0,  No pairing correlations (Hartree-Fock calculation),
\item If {\tt IPPFORCE}= 1,  Calculation for the density-dependent delta pairing force,
\item If {\tt IPPFORCE}= 2,  Calculation for the density-independent delta pairing force.
\end{itemize}
\item {\bf (l)} The quadrupole-constraint control parameter {\tt ICSTR}. If {\tt ICSTR}=0, the quadrupole constraint is not included, and the next two fields (m) and (n) are not used. If  {\tt ICSTR}=1, then:
\begin{itemize}
\item {\bf (m)} Constrained value of the quadrupole deformation $\bar{\beta}$.
The value of $\bar{\beta}$ defines the constrained quadrupole moment $\bar{Q}$ in Eq.\ (\ref{conn})
through:  ${\textstyle{\bar{Q}=\sqrt{\frac{5}{\pi}} <r^2>\bar{\beta}}}$. \tbd{what is $<r^2>$}
\item {\bf (n)} Parameter $\eta$ defining the stiffness $C_Q$
of the quadratic quadrupole constraint constant  by $C_Q=\eta (41 A^{-1/3})/(8Ab_0^2<r^2>)$.
\end{itemize}
\item {\bf  (o)} The number of gauge-angle points $L$ used for the particle number projection.
                 Note that the code always performs the PNP, even if pairing correlations are
                 not included.
\item {\bf  (p)} The particle-number-shift control parameter {\tt ISHIFT}. If, {\tt ISHIFT}=0, the particle-number-projection is performed on $N$ and $Z$, and the next two fields (q) and (r) are not used. If {\tt ISHIFT}=1, then:
\begin{itemize}
\item {\bf  (q)} Neutron-number shift {\tt KDN}, i.e., the projection is performed on neutron number $N$+{\tt KDN},
\item {\bf  (r)} Proton-number  shift {\tt KDZ}, i.e., the projection is performed on proton  number $Z$+{\tt KDZ}.
\end{itemize}
\item {\bf  (s)} Requested precision of convergence {\tt SI} (in
MeV). Iterations stop when changes of all mean-field and pairing matrix
elements between two consecutive iterations become smaller than the
value of {\tt SI}. Recommended value is $0.0001$.
\end{itemize}

\begin{table}
\caption[T1]{User-defined parameters of the Skyrme force, as given in the file {\tt forces.dat}.}
\label{tab2}
\vspace*{0.5cm}
\begin{center}
\begin{tabular}{|r|p{8cm}|}
\hline
Value & Description \\
\hline
    'SLY4'         & Skyrme-force acronym                              \\
     0             & Tensor term (0-excluded, 1-included)               \\
  $-$0.2488913d+04 & $t_0$                                              \\
     0.4868180d+03 & $t_1$                                              \\
  $-$0.5463950d+03 & $t_2$                                              \\
     0.1377700d+05 & $t_3$                                              \\
     0.8340000d0   & $x_0$                                              \\
  $-$0.3440000d0   & $x_1$                                              \\
  $-$1.0000000d0   & $x_2$                                              \\
     1.3540000d0   & $x_3$                                              \\
     0.1230000d+03 & $W_0$                                              \\
     6.0d0         & $1/\alpha$                                         \\
    20.735530d0    & $\hbar^2$/2m                                       \\
     0.160d0       & $\rho_0$       (saturation density for pairing)    \\
     1.0d0         & $\gamma$       (power of density for pairing)      \\
    60.0d0         & $\bar{e}_{\max }$ (pairing cut-off energy)         \\
     0.5d0         & $V_1$          (0-volume, 1-surface, 0.5-mixed)    \\
$-$244.7200d0      & $V_0$          (pairing strength)                  \\
\hline
\end{tabular}
\end{center}
\end{table}

After the solution is found, and if {\tt MAXI}$>$0, the code stores
files {\tt dnnn\_zzz.hel} (if the HO-basis run has been performed)
and/or {\tt dnnn\_zzz.tel} (if the THO-basis run has been performed).
Names of these files are automatically constructed based on
the input-data parameters {\tt ININ}, $N$, and $Z$, namely:

\begin{itemize}
\item {\tt d}   = `s', `p', or `o', for $|${\tt ININ}$|$ = 1, 2, or 3, respectively,
\item {\tt nnn} = three-digit value of $N$ with leading zeros included,
\item {\tt zzz} = three-digit value of $Z$ with leading zeros included.
\end{itemize}

These files can be used in a later run to restart calculations from
previously found solutions. For example, file {\tt s070\_050.tel} contains results of the THO-basis
calculation for $^{120}$Sn, which has been obtained by starting
from a spherical field. Note that the name of the file reflects
the starting deformation only, while it may, in fact, contain results
for another deformation that has been obtained during the iteration.

\subsection{Output Files}
\label{sec33}

The results are printed on the standard output file.
Each run produces a separate part of the output file; also
the HO run preceding a THO run produces one such part. Below
we briefly describe different sections of the output file.

\begin{itemize}
\item {\it Header}.
Contains the version number of the code, date
and time of execution, name of the element, and its particle,
neutron, and proton numbers.
\item {\it Input data}.
Contains a short summary of the input data for the requested run.
\item {\it Force}.
Lists the acronym and parameters of the Skyrme force, as well as parameters
of the pairing force.
\item {\it Numerical}.
Contains some information on numerical parameters and options used for the
given run.
\item {\it Regime}.
Gives the regime in which the code is run.
\item {\it Iterations}.
Shows brief information about iterations performed. One line of the
output file per each iteration is printed and contains the following columns:
\begin{itemize}
\item Iteration number {\tt i}.
\item Accuracy {\tt si}.
\item Current mixing parameter between the previous and current fields {\tt mix}.
\item Quadrupole deformation {\tt beta}, ${\textstyle{\beta=\sqrt{\frac{\pi}{5}}\frac{<\hat{Q}>}{<r^2>}}}$, for $\hat{Q}$ given in Eq.\ (\ref{QQQ}).
\item Total energy {\tt Etot}.
\item Particle number {\tt A}.
\item Neutron rms radius {\tt rn}.
\item Proton rms radius {\tt rp}.
\item Neutron pairing energy {\tt En}.
\item Neutron pairing gap {\tt Dn}.
\item Proton pairing energy {\tt Ep}.
\item Proton pairing gap {\tt Dp}.
\item Neutron Fermi energy {\tt Ln}.
\item Proton Fermi energy {\tt Lp}.
\end{itemize}
\item {\it Files}.
Contains information on the {\tt dnnn\_zzz.hel} or {\tt dnnn\_zzz.tel}
file written.
\item {\it Observables}.
Lists values of various observables calculated for the HFB state
without PNP and with the Lipkin-Nogami corrections, and then
those calculated for the PNP HFB state.
\end{itemize}

The same information, plus more results on the quasiparticle and
canonical states, is also stored in the file {\tt thoout.dat}.
However, this file is rewound after each run, so it contains
results of only the last run executed for the given input data file.

Files {\tt hodef.dat} and {\tt thodef.dat} contain synthetic results
of all runs, printed in the form of a single line per each performed
run. If the given run performs only an HO-basis calculation, or only
a THO-basis calculation, then only an entry in file {\tt hodef.dat}
or {\tt thodef.dat} is produced, respectively. On the other hand,
runs that perform both HO and THO calculations produce entries in
both these files. Lines in the files {\tt hodef.dat} and {\tt
thodef.dat} contain 105 columns each, and each column is described by
a name printed in the first header line. The names are
self-explanatory, and most often they correspond to the names used in
the present write-up. Names preceded by {\tt U:} pertain to results
obtained for the HFB states before PNP, while those beginning with
{\tt L} pertain to the results containing the Lipkin-Nogami
corrections. Names ending with {\tt t}, {\tt n}, or {\tt p} give
total, neutron, or proton observables, respectively.

\section{Conclusions}
\label{sec4}

The  code HFBTHO {\codeversion} is a tool of choice for
self-consistent calculations for a large number of even-even nuclei.
Several  examples of
deformed HFBTHO calculations,  recently implemented on parallel
computers, are  given in Ref.~\cite{[Sto03]}.
By creating a simple load-balancing routine that allows one to scale the
problem to 200 processors, it was possible to calculate the entire deformed
even-even nuclear mass table in a single 24 wall-clock hour run (or approximately
4,800 processor hours).

The crucial input for such calculations, which determines
the quality of results,  is the nuclear energy density
functional.
The development of  the ``universal"
nuclear energy density functional still remains
 one of the major challenges for nuclear theory.
While self-consistent  HFB methods
have  already
achieved a level of sophistication and precision which allows analyses
of experimental data for a wide range of properties and for arbitrarily
heavy nuclei (see, e.g., Refs. \cite{[Gor02],[Sam02],[Gor03]} for deformed
HFB mass table),
much work  remains to be done. Developing a universal nuclear density
functional  will require a better
understanding of the    density dependence, isospin effects,
and pairing,
as well as an improved treatment of symmetry-breaking effects and
many-body correlations.

In addition to systematic improvements of  the nuclear energy density
functional, there are several anticipated extensions of HFBTHO itself.
The  future enhancements  to  HFBTHO
will include the implementation of the full particle-number projection
{\it before} variation, extension  of code to odd particle numbers,
implementation of non-standard spin-orbit term and two-body center-of-mass
correction,
and evaluation  of dynamical corrections representing  correlations beyond
the mean field.

\bigskip

This work was supported in part by the U.S.\ Department of Energy
under Contract Nos.\ DE-FG02-96ER40963 (University of Tennessee),
DE-AC05-00OR22725 with UT-Battelle, LLC (Oak Ridge National
Laboratory), and DE-FG05-87ER40361 (Joint Institute for Heavy Ion
Research);  by the National Nuclear Security Administration under the
Stewardship Science Academic Alliances program through DOE Research
Grant DE-FG03-03NA00083;  by the Polish Committee for Scientific
Research (KBN); and  by the Foundation for Polish Science (FNP).


\end{document}